\documentclass[
preprint,
superscriptaddress,
amssymb,
amsmath
]{revtex4-1}

\usepackage{graphicx}% Include figure files
\usepackage{dcolumn}% Align table columns on decimal point
\usepackage{bm}% bold math
\usepackage{bm, color}
\usepackage{lineno}
\begin{document}
\title{Spin-orbitronic materials with record spin-charge conversion from high-throughput ab initio calculations}

\author{Yang Zhang}
\thanks{These two authors contributed equally}
\affiliation{Max Planck Institute for Chemical Physics of Solids, 01187 Dresden, Germany}
\affiliation{Leibniz Institute for Solid State and Materials Research, 01069 Dresden, Germany}
\author{Qiunan Xu}
\thanks{These two authors contributed equally}
\affiliation{Max Planck Institute for Chemical Physics of Solids, 01187 Dresden, Germany}
\author{Klaus Koepernik}
\affiliation{Leibniz Institute for Solid State and Materials Research, 01069 Dresden, Germany}
\author{Jakub \v{Z}elezn\'{y}}
\affiliation{Institute of Physics, Czech Academy of Sciences, Cukrovarnicka 10, 162 00 Praha 6 Czech Republic}
\author{Tom\'{a}\v{s} Jungwirth}
\affiliation{Institute of Physics, Czech Academy of Sciences, Cukrovarnicka 10, 162 00 Praha 6 Czech Republic}
\affiliation{School of Physics and Astronomy, University of Nottingham, NG7 2RD, Nottingham, United Kingdom}
\author{Claudia Felser}
\affiliation{Max Planck Institute for Chemical Physics of Solids, 01187 Dresden, Germany}
\affiliation{Center for Nanoscale Systems,Faculty of Arts and Sciences, Harvard University,11 Oxford Street, LISE 308 Cambridge, MA 02138, USA}
\author{Jeroen van den Brink}
\affiliation{Leibniz Institute for Solid State and Materials Research, 01069 Dresden, Germany}
\author{Yan Sun}
\email{ysun@cpfs.mpg.de}
\affiliation{Max Planck Institute for Chemical Physics of Solids, 01187 Dresden, Germany}

\begin{abstract}
Today's limitations of charge-based electronics in terms of power consumption
motivates the field of spintronics~\cite{Wolf2001}. 
The spin Hall effect (SHE) ~\cite{Dyakonov1971a, Dyakonov1971b, Sinova2015} is an important spintronics phenomenon, 
which allows transforming a charge current into a spin current and vice 
versa~\cite{Saitoh2006,Valenzuela2006,Hui_Zhao2006} without the use of magnetic materials or magnetic fields.
To gain new insight into the physics of the SHE and to identify materials 
with a substantial spin Hall conductivities (SHC), we performed high-precision, high-throughput {\it ab initio} electronic structure
calculations of the intrinsic SHC for over 20,000 non-magnetic crystals.  
The calculations reveal a strong and unexpected relation of the magnitude of the 
SHC with the crystalline symmetry, which we show exists because large SHC is 
typically associated with mirror symmetry protected nodal lines in the band structure.
From the new developed database,
we identify new promising materials. This includes
eleven materials with a SHC comparable or even larger than that the up to now record Pt as well as 
materials with different types of spin currents, which could allow for 
new types of spin-obitronics devices.

\end{abstract}

\maketitle

Even if the extrinsic spin Hall effect (SHE) was predicted almost 50 years
ago~\cite{Dyakonov1971a, Dyakonov1971b}, the SHE did not receive extensive
attention until last decade, after  theoretical studies of its intrinsic
mechanism~\cite{Murakami2003, Sinova2004} and its experimental observation~\cite{Kato2004,Wunderlich2005,Day2005}.
%The study of the quantized version of SHE ~\cite{Kane2005b,Kane2005,bernevig2006d,koenig2007,Hasan:2010ku, Qi2011RMP}
%and spin current from spin-momentum locked topological surface
%states~\cite{mellnik2014spin,khang2018conductive,fan2014magnetization,mahendra2018room}
%makes new approaches for the SHE.
The SHE causes an electrical current to generate a
transverse spin current ~\cite{Dyakonov1971a, Dyakonov1971b, Sinova2015}.
Vice versa, a spin current can also generate a transverse voltage drop by the
inverse SHE~\cite{Saitoh2006, Valenzuela2006,Hui_Zhao2006}.
Strong SHE materials being of central importance for the detection,
generation and manipulation of spin currents suggests performing a large-scale
screening to identify the most suitable materials for spintronics devices.
Experimentally, however such a large-scale screening is very impractical, as quantitative
determination of the spin Hall conductivity (SHC) by electrical measurement requires integrating
each material separately into a complex multicomponent microscopic transport
device~\cite{Saitoh2006,Valenzuela2006,Hui_Zhao2006}. Furthermore, different 
experiments often give very different results for the SHC since the spin current cannot be 
measured directly and other effects such as the inverse spin-galvanic effect complicate the 
interpretation of experiments.
Theoretically and computationally the situation is in principle much more straight-forward. 
An additional advantage of high-throughput calculations is that they can reveal 
further insight into the physics of the SHE and suggest general guidelines for designing new SHE materials.

In general, the SHE has two origins: an extrinsic contribution from scattering and an intrinsic
contribution from the electronic band structure. In this work we focus only on the intrinsic 
contribution for two reasons. First, the intrinsic contribution is typically the dominant 
one in systems with strong spin-orbit coupling \cite{Sinova2015}. Second, it can be 
accurately predicted theoretically as long as the electronic structure calculation is precise
enough~\cite{Murakami2003, Sinova2004,Guo2005, Xiao2010, Sinova2015} whereas the extrinsic 
contributions are much harder to calculate and strongly depend on the type of scattering and 
other parameters such as temperature. In many cases, the calculated intrinsic SHE yields good qualitative agreement
with experimental measurements~\cite{Sinova2015,Axel2013}. For example, 
the large intrinsic SHC of Pt ~\cite{Kimura2007, Takeshi2008,Miren2015} or 
the predicted sign change of SHC from Pt to Ta were experimentally
observed~\cite{Tanaka2008,Hahn2013}. Therefore, this database can be helpful for selecting 
new spin-to-charge conversion materials even if it cannot be expected that the predicted 
values will be precisely reproduced in experiments.
Apart from the SHC, there are other parameters that determine the 
usefulness of a material for spin-to-charge conversion. This is in particular the ratio 
of spin and charge conductivities, known as the spin Hall angle but also other parameters 
such as the spin diffusion length. We do not attempt to evaluate these parameters here, 
however we note that the charge conductivity is straightforward to measure experimentally 
and thus the spin Hall angles can be easily obtained from our calculations for materials 
in which the conductivity is experimentally known.

To perform the high-throughput simulations, we developed the automatic Wannier
function generating code,
which enables us to do high-throughput calculations of the intrinsic SHE for over 20,000
different known non-magnetic materials with a workflow that is shown in Fig. 1. We
first consider all lattice structures from the ICSD database~\cite{Mariette2004} 
and Materials Project~\cite{Anubhav2013,Ong2013}. Most of the studied materials 
are in the ICSD database (17,682 in total), which contains real, experimentally 
characterized materials. In addition we also consider computational materials
(i.e., materials that have not been experimentally synthesized, but have been 
predicted to be stable by \emph{ab-initio} calculations) from the Materials 
Project~\cite{Anubhav2013,Ong2013} since they are extracted from alloys or 
similar structures on the basis of experimentally known materials and can 
thus be possibly synthesized. Considering the limited accuracy of the density
functional theory (DFT) for the strongly correlated systems, we leave them out of
consideration (see Fig.1). In total we considered 17,682 real materials from the 
ICSD database and 2,486 calculational materials from the Materials Project.

These lattice structures are loaded into the full-potential local-orbital minimum-basis DFT
code of FPLO for the DFT calculations~\cite{Koepernik1999, perdew1996}.  After excluding the magnetic materials,
self-consistent nonmagnetic DFT calculations were performed. We
then project the Bloch wavefuctions into highly symmetric atomic-orbital-like Wannier functions,
and generate the corresponding tight binding model Hamiltonians. We note that the commonly
used approach based on maximally localized Wannier functions cannot easily be used for the
high-throughput calculations since it is not easily automatized and the resulting tight
binding Hamiltonians often contain spurious symmetry breaking. In our approach, 
we make use of the fact that FPLO code uses a local basis set, which makes it straightforward 
to generate a very high precision tight-binding Hamiltonian. Based on the tight binding
model Hamiltonian that fulfills all symmetries, the intrinsic spin Hall conductivities (SHCs) are
computed by the Kubo formula approach in the clean limit~\cite{Xiao2010,Sinova2004,Guo2005}.
To confirm the $k$-point convergence, a dense $k$-grid of $250\times250\times250$ was applied in
the SHC calculations. The accuracy of calculated SHC was double checked by the symmetry
analysis~\cite{Kleiner1966,Seemann2015}.  All of the SHC calculations are based on nonmagnetic
DFT calculations. Some of the considered materials may be magnetic, however, and for these the
SHC calculation would not be accurate (though it could in some cases correspond to temperatures
above the Curie temperature). We thus checked the magnetic states of materials with large SHC
($>$500 $(\hbar/e)(S/cm)$) and left the magnetic materials out of consideration. 
For materials that are in the Materials Project we have utilized
the information on the magnetic state from the Materials Project. For materials that are only
in the ICSD database, we have performed a collinear spin polarized DFT calculation, and classified
systems with a total moment larger than 0.05 $\mu_{B}$ per unit cell as magnetic.

The distribution of calculated SHC for the full set of materials is shown in Fig. 2 (details for
each separate compound are listed in the Supplementary material). From the more than 20,000
materials  we find only 1048 with a SHC larger than 500 $(\hbar/e)(S/cm)$, and only 169 above
1000 $(\hbar/e)(S/cm)$. From the distribution of SHC in the space groups (Fig. 2(a)), one sees
that the large value ($>$500 $(\hbar/e)(S/cm)$) is mainly contained in six blocks, which are
No. 61-63, No. 138-140, No. 160-164, No. 186-187, No. 191-194, and No. 221-229. To identify the
common characteristics of these space groups in the six blocks, one can directly check their
common symmetry operations. It turns out that all these space groups contain more than one
mirror plane. Further statistical analysis finds that materials with much larger SHC than 
the average have significantly more mirrors than materials with much smaller SHC than the 
average, see Fig. 2(b). This shows that surprisingly there is a strong connection between 
the magnitude of the SHE and crystalline symmetry. As we discuss below, this connection 
exists because of special features in the band structure, which are protected by mirror symmetry.

Chemical element analysis of the materials with a value of the SHC larger than 500 $(\hbar/e)(S/cm)$
shows that the $5d$ transition metals Pt, Ir, Hf, and Ta are the most common elements (see Supplementary material).
This is in accordance with the general intuition that to achieve a large SHC, on one hand, a
large spin orbit coupling is required and, on the other hand, rather extended electronic orbitals
should contribute to the electron transport. In these two respects the $5d$ transition metal
compounds are better than $4d$ and $3d$ ones. Indeed previous SHE studies confirm that materials
with large SHC are $5d$ transition metal related compounds~\cite{Sinova2015,Saitoh2006,Kimura2007,Tanaka2008,Hoffmann2013,Sahin2015}.

As highlighted in Fig. 2(a)  we identify 
eleven materials with a SHC comparable or even larger than Pt:
IrN, Bi$_2$In$_5$, Tl$_3$Ru, Pt$_3$Rh, CuPt$_7$, LiPt$_7$, Bi$_2$OsAu, Bi$_3$In$_4$Pb, HgOsPb$_2$, LiIr, and PtRh$_3$.
The compound with the largest SHC is IrN, which has SHC over 2900 $(\hbar/e)(S/cm)$, however,
we note that this is for a theoretically proposed phase of IrN.\cite{Li2012,Rached2011} The next largest is the
In$_5$Bi$_3$, which, unlike the other record compounds that all contain $5d$ elements,
is a pure $p$-orbital metal with a SHC above 2500 $(\hbar/e)(S/cm)$.
As we show in the following, the giant SHC in In$_5$Bi$_3$ originates
not only from its large spin orbit coupling but also from a set of special features in its band
structure. The fact that out of the 20,000 studied crystals, only a handful has 
SHC larger than Pt and none has a significantly higher value suggests that we may have approached 
the realistic limit of the intrinsic SHC in our calculations.

To analyse the origin of large SHC we consider the spin Berry curvature distribution in the Brillouin zone (in analogy
to the ordinary Berry curvature in the anomalous Hall effect ~\cite{Xiao2010}) of three selected materials
with different point groups: YIr, In$_5$Bi$_3$, and Pt.  Fig. 3(a-c) show the example of
YIr with space group $Pm\bar{3}m$ (No. 221) and point group
$m\bar{3}m$~\cite{Blazina1989}. This material has a simple electronic structure and it is, 
therefore, useful for illustrating the origin of the SHC. The large intrinsic SHC in this material is 
associated with nodal lines in the band structure as  illustrated in Fig. \ref{fig:group}(c). The nodal 
lines are one dimensional band crossings, which are in absence of the spin-orbit coupling protected by mirror 
symmetry. With spin-orbit coupling, the nodal lines are split, which gives rise to a large spin Berry curvature. 
It can be seen in Fig. \ref{fig:sample}(a) that the main contribution to the SHC in YIr is indeed around the nodal 
lines. The band structure forms an independent nodal ring in the $k_z$=$\pi$ plane, centered at
the $(\pi,0,\pi)$ point, which is protected by the mirror symmetry $M_z$. In combination with $c_4$ rotation 
symmetry with respect to the $x$, $y$,
and $z$ axes, there are 24 nodal rings in total in the cubic Brillouin zone (Fig. 3(a)). Because of the large spin Berry
curvature from the 24 nodal rings, the SHC of YIr reaches $\sim{1600}$ $(\hbar/e)(S/cm)$, see Fig. 3(c).

Apart from the record breaking In$_5$Bi$_3$, the high-throughput calculations identify four other
non-transition metal compounds (InBi, In$_2$Bi, In$_5$Bi$_2$Pb, and element Tl) with a large SHE.
The giant SHC in In$_5$Bi$_3$ is related to the mirror planes $m_{100}$ and $m_{110}$ which cause
two independent nodal lines in $k_x$=0 and $k_x+k_y$=0 planes, respectively.
In combination with the $m_{001}$ and $c_{4}^{z}$ rotation symmetry, there are 16 nodal lines in total.
In addition to these mirror symmetry protected nodal lines, In$_5$Bi$_3$ also contains special $PT$
symmetry protected nodal rings out of high symmetry planes. It is clear from Fig. 3(d) that the hot
lines of spin Berry curvatures are dominated by these nodal lines.

Performing a similar symmetry analysis on the electronic structure of Pt indicates that also in this
well-known SHE material the large SHC originates from mirror symmetry: it protects nodal rings whose
spin Berry curvature distribution is shown in Fig. 3(g). Interestingly, the contribution to the
spin Berry curvature of the high symmetry points of $L$ and $X$ are less than 15\% of the contribution
from the nodal lines. Even if the Berry curvatures themselves are very large around the two high symmetry
points of $L$ and $X$, their volumes are much smaller compared to the nodal lines. This brings to 
an important advantage of nodal rings in the context of the SHE: their significant dispersion in
energy space offers a large possibility to cross the Fermi level and contribute strongly to the SHC.

In all of the materials where we have explored the origin of large SHE in depth, 
we have found that the SHE mainly originates from symmetry protected nodal lines. This is the 
likely explanation for the strong statistical relation of large SHE to symmetry shown in 
Fig. \ref{fig:group}c. 
Importantly, similar relation to symmetry 
will likely exist also in other effects that have similar origin to the intrinsic SHE, such as 
the anomalous Hall effect or the anti-damping spin-orbit torque. Our result thus has quite 
fundamental significance: it shows that the symmetry determines not only the the presence or 
absence of the transport phenomena as has been previously thought, but can also have a strong 
influence on their magnitude.

In most materials, the symmetry of the SHE is such that the spin-polarization of the spin current
is perpendicular to both the spin current and the charge current. This is not a general rule,
however. In materials with low crystalline symmetry,
other types of spin current are allowed.\cite{Wimmer2015} This includes a spin current 
which flows in the transverse direction to the charge current like normal SHE but has spin-polarization 
along the spin current flow direction or longitudinal spin currents:  Spin currents which flow in the 
same direction as the charge current (which can then have either spin-polarization parallel with the 
flow direction or perpendicular to it), as illustrated in Fig. \ref{fig:spin_current}. These types of 
spin currents could allow for new functionalities in spintronic devices, however they have not received 
much experimental attention since materials where they are allowed are quite rare. A theoretical 
screening to identify promising materials is thus essential. We note that the SHE normally refers to 
a spin current flowing in transverse direction to the charge current and thus it is not clear whether 
the longitudinal spin currents should be referred to as the SHE. Nevertheless, we stress, that the 
longitudinal spin currents have the same origin as the conventional SHE.

SHE which flows in the transverse direction to the charge current and has a spin-polarization
 along the spin current flow direction  is of great interest for spin-orbit
torques in heavy metal / ferromagnet bilayer systems (or other similar systems)~\cite{MacNeill2016} (see Fig. \ref{fig:spin_current}b). In these
structures, the SHE from the heavy metal layer flows into the ferromagnetic layer and thus exerts
a torque on the magnetization ~\cite{Manchon2018}. For
scalability, it is preferential to utilize ferromagnet with a perpendicular magnetic anisotropy
(PMA). However, for deterministic field-free switching of the PMA systems, it is necessary to have a spin
current with spin-polarization perpendicular to the interface (and thus parallel to the spin current
flow direction). This is not allowed by symmetry in materials commonly used in these systems
such as Pt and thus materials with lower symmetry are needed. Our calculations reveal that a
large SHC with spin-polarization along spin current can occur, but is relatively rare. Out of the
1048 materials with SHC larger than 500 $(\hbar/e)(S/cm)$, only 2 have spin-polarization
along the spin current: Ni$_2$P$_6$W$_4$ and Ba$_2$C$_4$S$_4$N$_4$. We have identified 32 other
materials with SHC larger than 250 $(\hbar/e)(S/cm)$, which are listed in the Supplementary material.

The longitudinal spin currents have been experimentally studied in ferromagnetic systems, 
where the origin of such currents lie in the ferromagnetic order. It has been, however, predicted 
earlier that the same mechanism generating the  SHE can also generate such spin currents \cite{Wimmer2015}. 
In our database we identify a number of materials which exhibit large longitudinal spin currents, 
with spin-polarization parallel or perpendicular to the
spin-current flow, see Fig. 4 (c, d). The largest longitudinal spin current 
can reach $\sim$610 $(\hbar/e)(S/cm)$ at $\sigma_{zz}^{z}$
in P$_7$Ru$_{12}$Sc$_2$. We list all the materials with longitudinal SHC larger than 
$250 (\hbar/e)(S/cm)$ in the supplementary material. The longitudinal spin current 
in non-magnetic crystals may offer a new platform
for the study and utilization of the spin current in non-magnetic materials. 

Our calculations reveal that the origin of large SHC is usually 
associated with mirror-symmetry protected nodal lines in the band structure, 
which results in a strong correlation between the crystalline symmetry and the SHC 
magnitude. This suggests that for the design of new SHC materials it is beneficial to 
consider high-symmetry materials with a large number of mirror planes. In addition we 
find that, apart from the obvious requirement of the presence of heavy elements with a
strong spin-orbit coupling, the presence of $5d$ transition metal elements
is advantageous, but not decisive. We identify a number of promising spin-to-charge 
conversion materials, including 169 materials with SHC
above 1000 $(\hbar/e)(S/cm)$, 11 materials with SHC comparable or even larger
than the up to now record 
Pt and materials in which the symmetry of SHE is lower thus allowing for 
different types of spin currents. 
With these general design principles on one hand, and the specific information on each separate compound on the
other hand, our high-throughput database can provide a powerful tool for the experimental design of spintronic
devices.
%%%%%%%%%%%%%%%%%%%%%%%%%%%%%%%%%%%%%%%%%%%%%%%%%%%%%%%%%%%%%%%%%%%%%%%%%

\textbf{Method:}The $ab-initio$ calculations were performed based one density functional
theory (DFT) by using the FPLO code~\cite{Koepernik1999}.
The exchange and correlation energies
were considered in the generalized gradient approximation (GGA)~\cite{perdew1996}.
The $k$-point grid in DFT calculation is setted as $12 \times 12 \times 12$,
and the criterion of total energy convergency is bellow 10$^{-6}$ eV.
To perform the SHC, we have projected
the Bloch wavefunctions to atomic-orbital-like Wannier functions by an
automatically procedure. To make the Wannier projection automatically
and accurately, we set the constraint
of the mean error between DFT and tight binding bellow 0.02 eV in the energy
window of E$_f$-2.0 eV to E$_f$+2.0 eV. The symmetry of the Wannier functions
is well restored from the FPLO DFT code, where the
Wannier functions are directly mapped from the atomic orbitals without maximum
localization process.
Starting from the tight binding model
Hamiltonian, the SHC was calculated by the linear response Kubo
formula approach in the clean limit~\cite{Xiao2010, Sinova2015}:
%\begin{widetext}
\begin{equation}
\begin{aligned}
\sigma_{ij}^{k}={e} \int_{_{BZ}}\frac{d\vec{k}}{(2\pi)^{3}}\underset{n}{\sum}f_{n\vec{k}}\Omega_{n,ij}^{S,k}(\vec{k}), \\
\Omega_{n,ij}^{S,k}(\vec{k})=-2Im\underset{n'\neq n}{\sum}\frac{\langle n\vec{k}| J_{i}^{k}|n'\vec{k} \rangle \langle n'\vec{k}| v_{j}|n\vec{k}\rangle}{(E_{n\vec{k}}-E_{n'\vec{k}})^{2}}
\end{aligned}
\label{SHC}
\end{equation}
%\end{widetext}
where $f_{n\vec{k}}$ is the Fermi--Dirac distribution for the $n$-th band.
$J_{i}^{k}=\frac{1}{2}\left\{ \begin{array}{cc}
{v_{i}}, & {s_{k}}\end{array}\right\} $
is the spin current operator with spin operator ${s}$, velocity operator
${v_{i}}$, and $i,j,k=x,y,z$.
$| n\vec{k} \rangle$ is the eigenvector for the Hamiltonian
${H}$ at eigenvalue $E_{n\vec{k}}$.
$\Omega_{n, ij}^{S,k}(\vec{k})$
is referred to as the spin Berry curvature for the n-th band at point $\vec{k}$
as an analogy to the ordinary Berry curvature.

\begin{acknowledgments}
This work was financially supported by the ERC Advanced Grant No.
291472 `Idea Heusler', ERC Advanced Grant No. 742068--TOPMAT,
Deutsche Forschungsgemeinschaft DFG under SFB 1143, 
EU FET Open RIA Grant No. 766566 grant (ASPIN),
the ministry of Education of the Czech Republic Grants No. LM2018110,
and LNSM-LNSpin and the Grant Agency of the Czech Republic Grant No. 19-18623Y
This work was performed in part at the Center for Nanoscale Systems (CNS),
a member of the National Nanotechnology Coordinated Infrastructure Network
(NNCI), which is supported by the National Science Foundation under NSF
award no. 1541959. CNS is part of Harvard University. Most of our calculations
are carried on Cobra cluster of MPCDF, Max Planck society.
\end{acknowledgments}

%%%%%%%%%%%%%%%%%%%%%%%%%%%%%%%%%%%%%%%%%%%%%%%%%%%%
\bibliographystyle{ieeetr}
\bibliography{transport}

% Figure 1 %%%%%%%%%%%%%%%%%%%%%%%%%%%%%%%%%%%%%%%%%%%%%%%%%%%%%%%%%%%%%%%%%%%%%
\begin{figure*}[htb]
\centering
\includegraphics[width=0.9\textwidth]{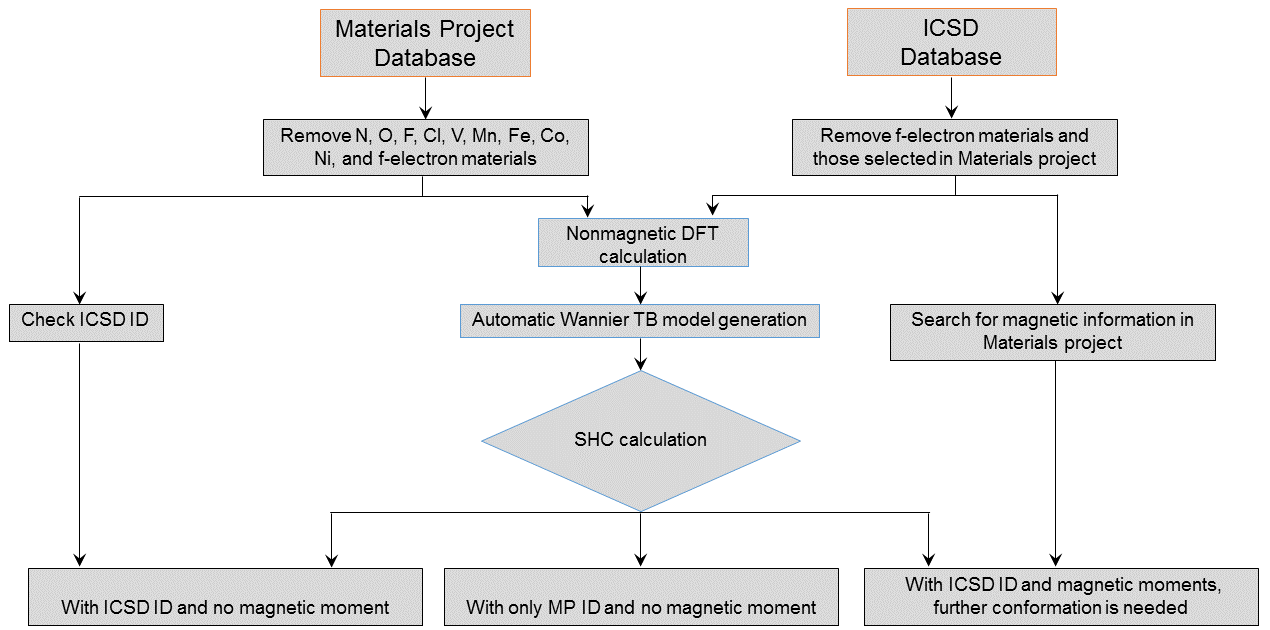}
   \caption{
%{\color{red} {\bf JvdB: please change "f materials" into "f-electron materials"}}
Workflow of the high-throughput calculations for intrinsic SHC.
The lattice structures are loaded from the Materials
Project and ICSD database. Since the DFT is not accurate for the strong correlated system,
all the compounds consisted from magnetic atoms V, Mn, Fe, Co, Ni or elements of N, O, F, and Cl 
are removed for Materials Project structures. All the f-electron related materials of both
databases are excluded. The known Mott insulators to our knowledge are also removed. For the repeated
lattice structures from the International Crystal Structure Database
(ICSD) ~\cite{Mariette2004} and Materials Project, we only calculate one
single structure. With this input list we performed self-consistent calculations with 
inclusion of spin orbit coupling for all compounds. The Bloch wavefunctions were 
automatically projected into highly symmetric atomic-orbital-like Wannier functions and 
the corresponding tight bonding model Hamiltonian was constructed. Based on the highly 
symmetric tight binding model Hamiltonians, the intrinsic SHC are computed
automatically based on Kubo formula at the clean limit.
}
\label{fig:workflow}
\end{figure*}
%%%%%%%%%%%%%%%%%%%%%%%%%%%%%%%%%%%%%%%%%%%%%%%%%%%%%%%%%%%%%%%%%%%%%%%%%%%%%%%%

% Figure 2 %%%%%%%%%%%%%%%%%%%%%%%%%%%%%%%%%%%%%%%%%%%%%%%%%%%%%%%%%%%%%%%%%%%%%
\begin{figure*}[htb]
\centering
\includegraphics[width=0.85\textwidth]{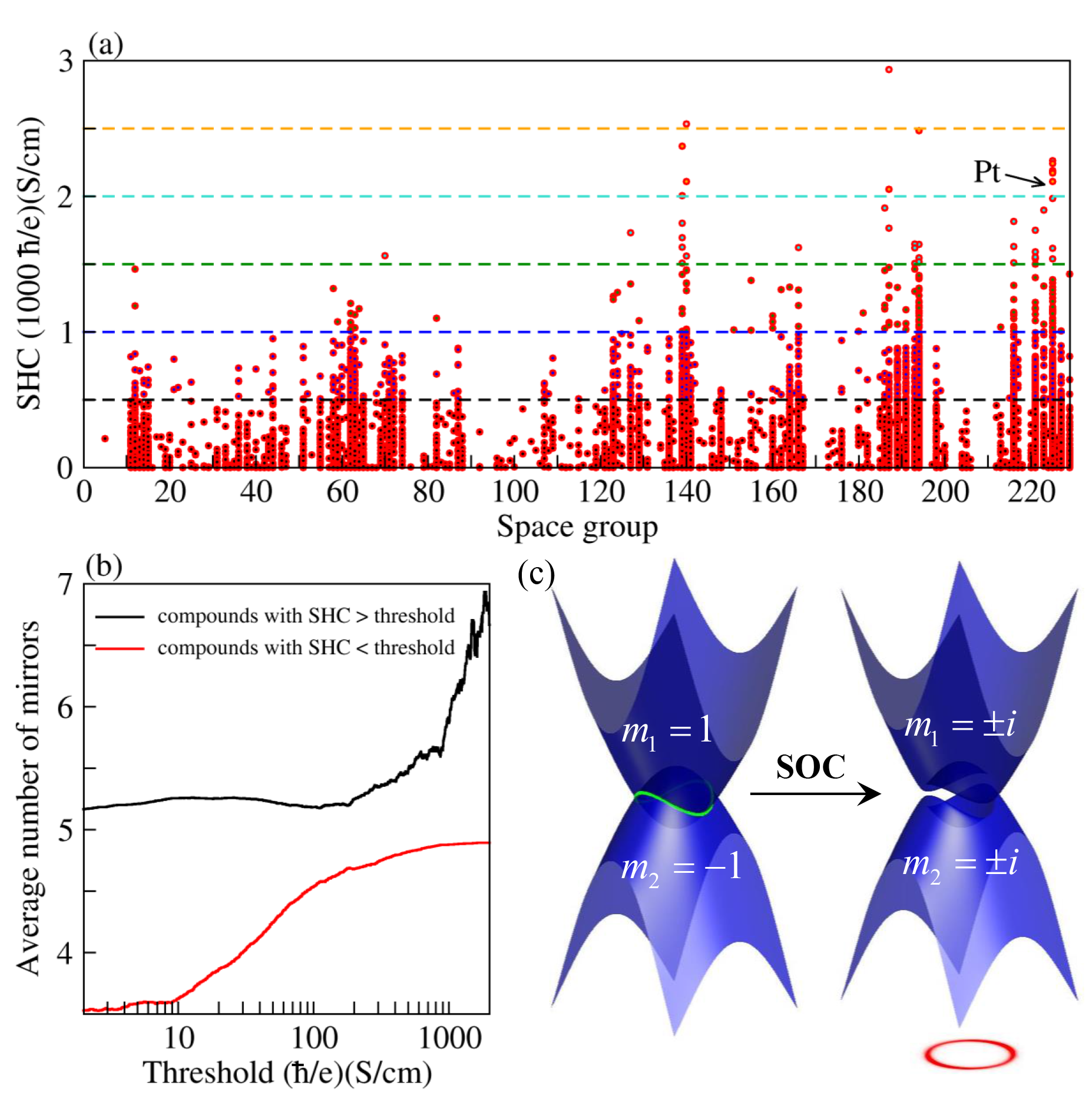}
   \caption{
Calculated intrinsic SHC.
(a) SHC distribution for compounds as a function of space group.
The SHC bellow 500, between 500-1000, between 1000-2000, between 2000-2500, and
above 2500 $(\hbar/e)(S/cm)$ are labeled by black, blue, green, orange, and
grey dots, respectively. 
(b) The average number of mirror symmetry operations in the space 
groups for materials with maximum value of SHC larger than threshold (black line) 
and smaller than threshold (red line) as a function of the threshold.
To determine the relation between symmetry and SHC, we only choose the maximum 
tensor element for each compound in the $3\times3\times3$ SHC tensor. We note that 
for the the statistical analysis we have also included the calculations for 
materials that are likely to be magnetic. (c) The illustration of the nodal 
line mechanism. Left figure shows the nodal line, which is present without 
spin-orbit coupling. The right figure shows the splitting of the bands due 
to spin-orbit coupling, which results in a large spin Berry curvature along 
the original nodal line, as illustrated by the red circle.
}
\label{fig:group}
\end{figure*}
%%%%%%%%%%%%%%%%%%%%%%%%%%%%%%%%%%%%%%%%%%%%%%%%%%%%%%%%%%%%%%%%%%%%%%%%%%%%%%%%

\begin{figure*}[htb]
\centering
\includegraphics[width=0.85\textwidth]{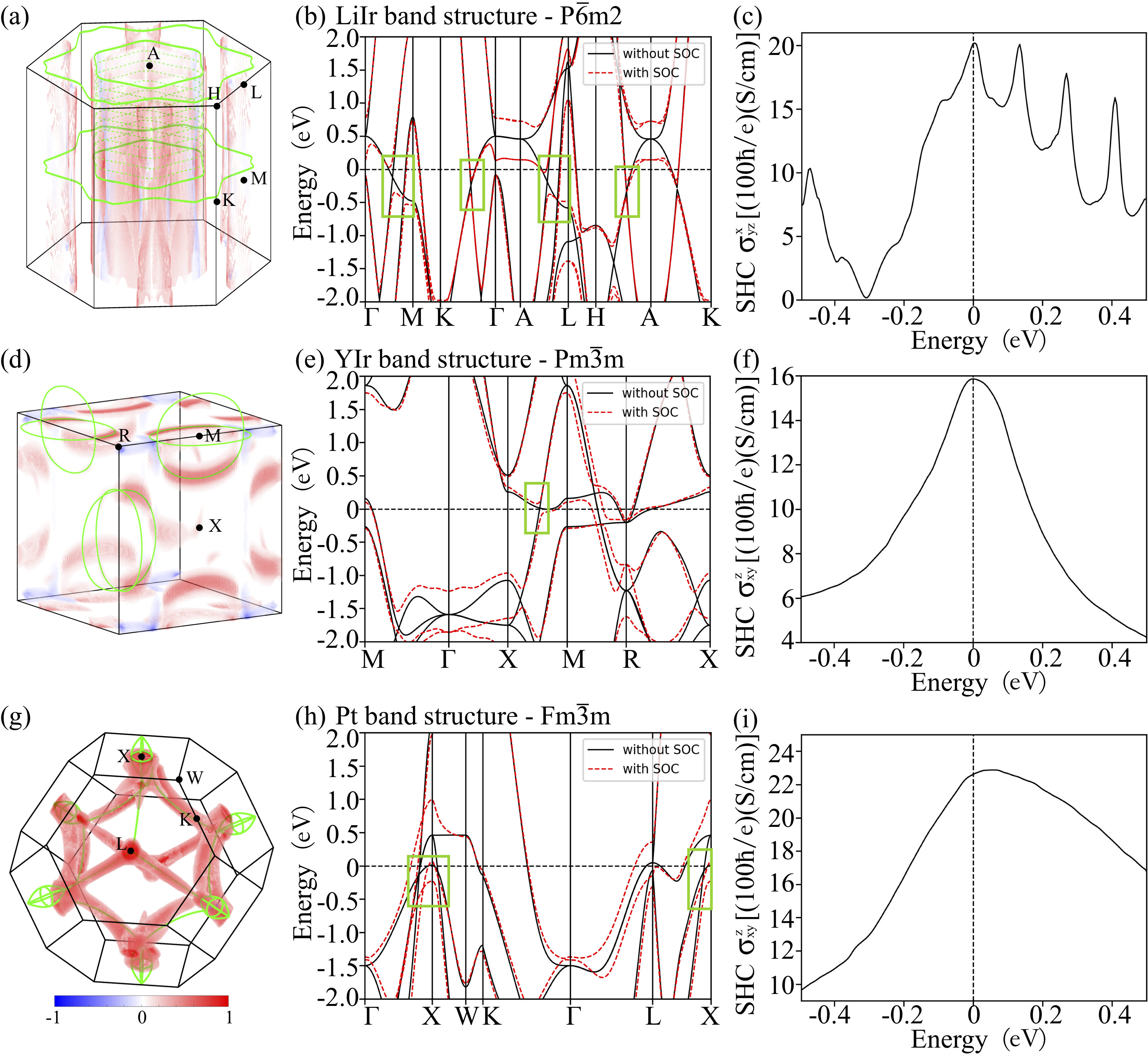}
   \caption{
Three typical examples with large SHC with different spaces.
Spin Berry curvature and nodal line distributions for (a) 
YIr with point group $m\bar{3}m$, (d) In$_5$Bi$_3$ with point group
$4/mmm$, and (g) Pt with point group $m\bar{3}m$, respectively.
The green lines represent nodal lines.
The color bar is in arbitrary unit. Energy dispersion for (b) YIr, 
(e) In$_5$Bi$_3$, and (h) Pt, respectively.
The position of nodal lines are labeled using green rectangles.
Fermi level dependent SHC for (c) YIr, (f) In$_5$Bi$_3$, and (i) Pt, respectively.
}
\label{fig:sample}
\end{figure*}
%%%%%%%%%%%%%%%%%%%%%%%%%%%%%%%%%%%%%%%%%%%%%%%%%%%%%%%%%%%%%%%%%%%%%%%%%%%%%%%%

\begin{figure*}[htb]
\centering
\includegraphics[width=0.85\textwidth]{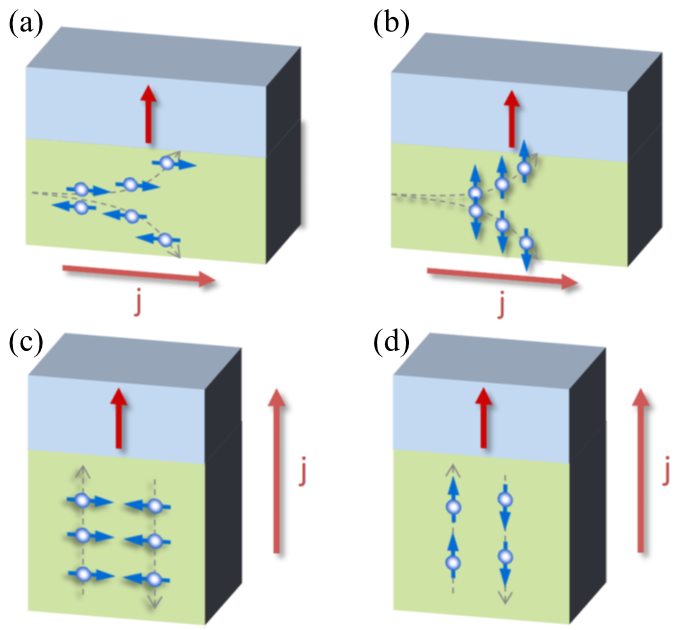}
   \caption{
The heterojunction of ferromagnet and non-magnets with 
four types of spin current obtained in this database. The top layer 
is ferromagnet and the bottom layer is non-magnet
(a) Traditional SHE with spin-polarization of the spin current
perpendicular to both the spin current and the charge current.
(b) Spin-polarization of the spin current along the direction of
the spin-current flow. (c) Longitudinal spin current with spin-current flow
along the electric field and perpendicular to spin-polarization.
(d) Longitudinal spin current with both of spin-polarization and spin-current 
flow along the electric field.
}
\label{fig:spin_current}
\end{figure*}
%%%%%%%%%%%%%%%%%%%%%%%%%%%%%%%%%%%%%%%%%%%%%%%%%%%%%%%%%%%%%%%%%%%%%%%%%%%%%%%%

\end{document}